\documentclass[10pt]{article}

\usepackage[utf8]{inputenc}
\usepackage[T1]{fontenc}
\usepackage{mathptmx}
\usepackage[scaled=0.92]{helvet}
\usepackage{amsmath,amssymb}
\usepackage{graphicx}
\usepackage[table]{xcolor}
\usepackage{booktabs}
\usepackage{array}
\usepackage{multirow}
\usepackage{caption}
\usepackage{enumitem}
\usepackage{titlesec}
\usepackage[numbers,sort]{natbib}
\usepackage[a4paper,top=2.45cm,bottom=2.45cm,left=2.3cm,right=2.3cm]{geometry}
\usepackage[hidelinks]{hyperref}
\usepackage{xurl}
\usepackage{placeins}

\definecolor{ncehead}{HTML}{27374D}
\definecolor{nceblue}{HTML}{3F6FA3}
\definecolor{ncepink}{HTML}{C75B70}
\definecolor{ncegreen}{HTML}{3E8E5E}
\definecolor{ncepurple}{HTML}{7E5AA2}
\definecolor{rowpastel}{HTML}{EAF2FB}
\definecolor{rowpastelb}{HTML}{F3EEF8}
\definecolor{headpastel}{HTML}{D7E6F4}

\titleformat{\section}{\normalfont\large\bfseries}{\thesection.}{0.5em}{}
\titleformat{\subsection}{\normalfont\normalsize\bfseries}{\thesubsection.}{0.5em}{}
\titlespacing*{\section}{0pt}{1.4ex plus 0.6ex}{0.7ex}
\titlespacing*{\subsection}{0pt}{1.1ex plus 0.5ex}{0.5ex}

\hypersetup{colorlinks=true,linkcolor=nceblue,citecolor=ncegreen,urlcolor=nceblue}

\begin{document}
\pagestyle{plain}

\begin{center}
{\LARGE\bfseries Design and Development of a Neuromorphic Silicon Suite: PVT Sensing, Stochastic LIF Inference, On-Chip STDP Learning, and Crossbar Programming}\\[10pt]
{\large Poornima Kumaresan and Santhosh Sivasubramani\textsuperscript{*}}\\[4pt]
{\itshape Intrinsic Lab, Centre for Sensors, Instrumentation and Cyber-Physical System Engineering (SeNSE), Indian Institute of Technology Delhi, New Delhi 110016, India}\\[3pt]
\textsuperscript{*}Corresponding author. E-mail: \href{mailto:ssivasub@iitd.ac.in}{ssivasub@iitd.ac.in}, \href{mailto:ragansanthosh@ieee.org}{ragansanthosh@ieee.org}
\end{center}

\vspace{0.4em}
\noindent\textbf{Abstract.}
Edge neuromorphic systems require compact, configurable hardware blocks that combine probabilistic inference, local learning, and an interface to emerging analogue memory. This paper presents the design of a suite of four interface-compatible digital intellectual property blocks implemented as standard-cell CMOS on the SkyWater 130\,nm process: a process, voltage and temperature (PVT) sensor built from five selectable ring oscillators that also provides a jitter-based true-random-number generator and a frequency-bounds health monitor; a stochastic leaky integrate-and-fire (LIF) neuron with a configurable linear-feedback shift register, a programmable activation lookup table, and a refractory period; an on-chip spike-timing-dependent plasticity (STDP) learning controller with a programmable plasticity curve and reward-modulated, eligibility-trace, and anti-Hebbian modes; and a controller for memristive crossbar arrays supporting forming, set, reset, read and automated current-voltage sweep with current-compliance limiting and half-select biasing. All four blocks share a common serial peripheral interface (SPI) register file, and the sensor additionally exposes a direct parallel readout. Each block occupies a single tile and targets a 50\,MHz system clock. Each block was verified with cocotb tests at register-transfer and gate level, 99 cases in total, all passing, and was taken through an open standard-cell flow and submitted for tapeout through the Tiny Tapeout shared-silicon programme. Behavioural simulation reproduces the expected membrane dynamics, threshold-controlled firing rate, and the asymmetric potentiation and depression characteristic. Mapped to the open standard-cell library, each block occupies a post-synthesis cell area of 9.3 to 10.6 thousand square micrometres and places at 61 to 70 per cent utilisation on a single tile, meets the 50\,MHz timing constraint with positive setup and hold margin after clock-tree synthesis, and draws an estimated 0.64 to 0.70\,mW under a default switching-activity assumption. The work provides a coherent set of building blocks for PVT-aware stochastic computation with local plasticity and crossbar programming, together with a register-based programming model whose parameters span the requirements reported for memristive and spintronic device arrays. All results are from simulation and the implementation flow; no fabricated silicon is reported.

\vspace{0.5em}

\vspace{0.2em}

\section{Introduction}
Battery-powered and embedded systems increasingly classify, detect, and adapt close to where their data are produced, under power budgets that clocked processors meet poorly once the workload includes continual learning. Neuromorphic hardware targets this regime by representing information as sparse events and by computing where the data already reside, which lowers both the arithmetic and the data movement that dominate energy use in conventional architectures \cite{davies2018loihi,hai2026a,khanday2026artifi}. In practice an event-driven learning system is assembled from a small set of primitives: a source of controlled randomness for probabilistic activation, a neuron with adjustable excitability and leak, a local rule that updates synaptic weights from spike timing, and a programmable interface to the device that stores those weights.

Each primitive has a mature literature. Large digital processors integrate programmable neurons and on-chip learning at scale \cite{davies2018loihi,arfa2025hardwa,matsuo2024unsupe}; compact spiking blocks have been built in standard cells and on field-programmable devices \cite{maachi2024effici,venkateswara2024effici,sharma2026highle}; stochastic activation is generated with linear-feedback shift registers \cite{lee2024design,saikia2024adapti,akter2025highen}; spike-timing-dependent plasticity has been demonstrated in many synaptic technologies \cite{mostafa2014a,daddinounou2024bisigm,castro2025spiket}; and resistive crossbars with their programming peripherals are an active area of device and circuit work \cite{youn2025ferroe,naqi2024large,seiler2025specif}. These results are almost always reported one primitive at a time, at different technology nodes, with interfaces, clocking, and verification chosen for the block in isolation. A practitioner who needs a complete sense, infer, learn, and store chain therefore collects blocks from different sources and reconciles their conventions afterwards.

That reconciliation is the practical bottleneck. Blocks designed independently rarely share a register model or a host protocol, so an integrator rewrites configuration logic, re-times asynchronous interfaces, and re-establishes a verification flow for each combination. Few of the primitives are available together on a single, openly documented node where the same tools take every block from register-transfer code to a manufacturable layout. The building blocks of neuromorphic computation exist, but, to the authors' knowledge, a coherent, interface-compatible, and verification-consistent set of them at one open node has not been reported.

This paper presents four such blocks designed together on the SkyWater 130\,nm node: a ring-oscillator process, voltage and temperature sensor; a stochastic leaky integrate-and-fire neuron; a spike-timing-dependent plasticity controller; and a controller for memristive crossbar arrays. All four blocks expose one serial register model, so a single host can configure every block through one access pattern, and each block also operates on its own. Every block is taken through the same flow, from register-transfer code and directed simulation to a design-rule-clean layout, with one verification methodology applied throughout. For an integrator this removes the per-block interface and verification rework: the sensor, the neuron, the learning controller, and the crossbar controller are all configured by the same access sequence, and the sensor characterises the operating corner that the host reads alongside the other blocks.

The specific contributions are the following. First, a register-transfer-level (RTL) description of the four blocks, with the design choices that keep each within a single tile and a 50\,MHz budget. Second, a shared serial interface and a uniform sixteen-entry register map, including the synchronisation and bus-isolation logic that a multi-block chip requires. Third, a directed verification campaign of 99 test cases at register-transfer and gate level, with design-rule and layout-versus-schematic sign-off and an automated layout flow. Fourth, behavioural characterisation from the register-transfer models, covering the realised stochastic activation and the input-output transfer of the neuron, the programmable plasticity curve, and the modelled sensor response. What is new is not an individual block but a compatible set built to one interface and one verification methodology at a single open node; the source, test benches, and layout configuration for every block are public and are linked where each block is described. In plain terms, the contribution is practical rather than a new device or algorithm: about a dozen established neuromorphic capabilities, spanning multi-function ring-oscillator sensing with on-chip entropy and aging detection, stochastic integrate-and-fire inference with a refractory period, pair-based and three-factor plasticity with an anti-Hebbian mode, and compliance-limited crossbar programming with half-select biasing and pulse trains, are packed into four single-tile blocks that share one control interface and one verification flow on a single openly documented process, so that they can be reused and combined with much less integration effort than assembling equivalent blocks from separate sources. The work does not claim a new circuit principle or a measured efficiency record; its value lies in the coherence, openness, and verification of the set. The individual functions, including reward-modulated and eligibility-trace plasticity, ring-oscillator entropy and aging detection, and compliance-limited crossbar programming, are established techniques that the suite implements and releases openly rather than introduces.

The remainder of the paper is organised as follows. Section~\ref{sec:bg} reviews related neuromorphic hardware. Section~\ref{sec:arch} describes the system architecture and the shared interface. Sections~\ref{sec:ro} to~\ref{sec:xbar} detail the four blocks. Section~\ref{sec:prog} presents the programming model. Section~\ref{sec:verif} describes verification, and Section~\ref{sec:impl} reports the physical implementation. Section~\ref{sec:disc} discusses trade-offs and the path to silicon measurement, and Section~\ref{sec:concl} concludes.

\section{Background and related work}\label{sec:bg}
\textbf{Neuromorphic processors and accelerators.} Digital neuromorphic processors such as Loihi established that programmable on-chip learning, configurable neuron models, and event-driven communication can be integrated at scale \cite{davies2018loihi}. Subsequent work has mapped recurrent and quantised spiking networks onto such platforms and has examined their behaviour under deployment constraints \cite{arfa2025hardwa,boeshertz2024accura,matsuo2024unsupe}. At a smaller scale, field-programmable and standard-cell implementations show that compact spiking blocks can be realised without full-custom analogue design \cite{maachi2024effici,venkateswara2024effici,sharma2026highle}.

\textbf{Neuron circuits.} The LIF model remains the most common neuron model in spiking hardware because it captures temporal integration with minimal state \cite{venkateswara2024effici}. Implementations range from analogue silicon neurons \cite{srivastava2016silico,orima2024bifurc} through emerging-device and memristive neurons \cite{priyanka2024emulat,gangu2026d,faizan2025a,neifar2025design,brand2026canoni} to fully digital datapaths that prioritise leakage and event-driven efficiency \cite{krausse2024explor}. Stochastic activation, in which a neuron fires probabilistically as a function of its input, reduces sensitivity to input noise and supports sampling-based computation \cite{wang2014a2}. Efficient pseudorandom generation is therefore a core requirement, and linear-feedback shift registers (LFSRs) are the standard primitive for stochastic computing because of their low area and high throughput \cite{lee2024design,saikia2024adapti,akter2025highen,ichikawa2025random}. The neuron in this work combines a configurable-polynomial LFSR with a programmable activation table and a saturating integrator. This digital, pseudorandom-driven neuron with a refractory period follows the model used in large neurosynaptic cores \cite{cassidy2013}, and recent stochastic-computing neurons use the same ingredients \cite{sclif2026}.

\textbf{Local learning.} STDP adjusts a synaptic weight according to the relative timing of pre- and post-synaptic spikes and is a widely used unsupervised rule in neuromorphic systems \cite{bipoo1998,mostafa2014a}. Hardware realisations have been reported for magnetic tunnel junctions, domain-wall devices, and electrochemical transistors \cite{daddinounou2024bisigm,castro2025spiket,peyal2025demons}. A digital controller that stores the plasticity curve in a lookup table decouples the learning rule from the synaptic medium, which is the approach taken here. Reward-modulated and three-factor variants, in which a global signal gates timing-driven updates through an eligibility trace, are well established \cite{izhikevich2007,florian2007,legenstein2008,fremaux2016}, and compact digital implementations have been reported \cite{quintana2022}.

\textbf{Crossbars and in-memory computing.} Resistive crossbars implement weighted summation in the analogue domain and are a widely studied candidate for synaptic storage \cite{youn2025ferroe,naqi2024large,wang2024reserv}, and complete compute-in-memory accelerators have been built on this principle \cite{kim2022neuroc,hokenmaier2025a}. Practical operation requires forming, set and reset pulses, read sensing, and current-voltage characterisation, together with addressing and timing control \cite{seiler2025specif,chowdhury2026inmemo,hassan2017hardwa,yakopcic2016memris}. Precise programming relies on current compliance and write-verify pulse trains \cite{alibart2012}, and half-select bias schemes mitigate sneak-path disturbance in the array \cite{cassuto2016}. Device studies continue to report new resistive materials and switching mechanisms with distinct programming requirements \cite{santra2025resist,bature2025analys,reddy2026chromi,mukherjee2026integr,kim2025halide}. A device-agnostic controller that exposes pulse width, voltage code, and sweep parameters through registers, as developed here, accommodates this diversity.

\textbf{Open physical design at 130\,nm.} The SkyWater 130\,nm process and its associated open process design kit, released with support from Google, together with the OpenLane and OpenROAD flows, have made standard-cell tape-out accessible for academic blocks \cite{shalan2020buildi,herman2023design}. A growing body of work characterises standard cells, memory, and analogue blocks on this node using these tools \cite{rahman2024standa,rahman2024timing,bhatt2025explor,montanares2025openso,jaramillotoral2025analog}. The present suite is implemented on the same node and flow, which fixes a stable, well-documented target for the four blocks. The blocks were submitted through the Tiny Tapeout shared-silicon programme, which provides a standardised tile and pin harness. Open neuromorphic hardware has begun to appear on this and similar flows, including open spiking accelerators on the SkyWater node \cite{modaresi2023}, open digital neuromorphic processors with on-chip learning \cite{frenkel2019,frenkel2022}, and open field-programmable emulators \cite{neurocorex2025}; these are individual cores, whereas the present work offers several primitives behind one register interface.

\section{System architecture}\label{sec:arch}

\subsection{Pipeline organisation}
\begin{figure}[tbp]
\centering
\includegraphics[width=0.92\textwidth]{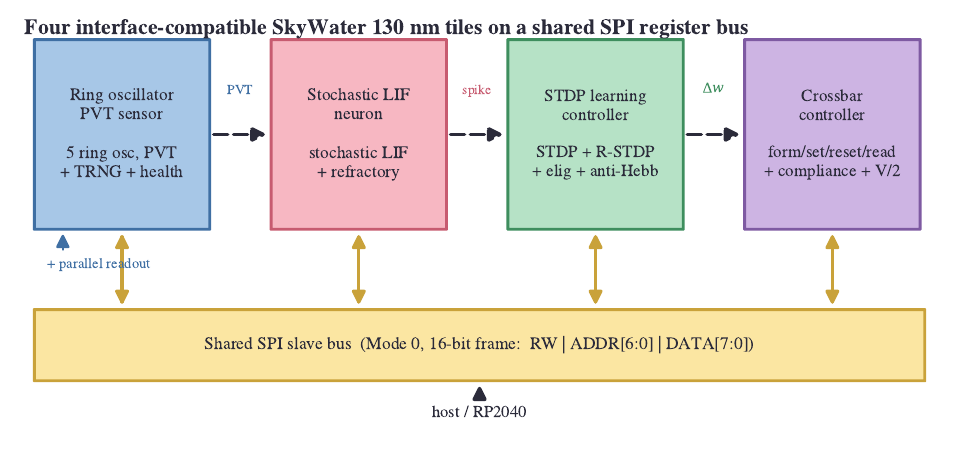}
\caption{Intended composition of the four neuromorphic blocks. The dashed arrows show the design-level dataflow for which the blocks were made interface-compatible: PVT characterisation feeding stochastic inference, spike-timing-dependent learning, and crossbar programming. All four blocks share a serial peripheral interface register bus; the ring-oscillator sensor additionally exposes a direct parallel readout. Each block is implemented and verified individually; the blocks are not co-simulated or connected in a single netlist in this work. The host controller, for example an RP2040 microcontroller, drives the shared bus.}
\label{fig:system}
\end{figure}

The suite is organised as the four-stage pipeline shown in figure~\ref{fig:system}, an intended dataflow across interface-compatible blocks that are verified individually rather than as one co-simulated netlist. The ring-oscillator sensor characterises the process corner and the supply and temperature conditions of the die, producing a frequency code that establishes a timing and bias baseline. The stochastic neuron performs probabilistic integrate-and-fire inference and emits a spike train. The STDP controller converts the relative timing of pre- and post-synaptic spikes into signed weight updates. The crossbar controller applies programming pulses to a resistive array to store the updated weights in the synaptic medium. Each block is independently testable and is mapped to a single Tiny Tapeout 1x1 tile of the target shuttle, about 161\,$\mu$m by 112\,$\mu$m.

Each block is also useful in isolation. The sensor is a self-contained PVT monitor, the neuron is a standalone stochastic spiking element, the learning controller is a reusable plasticity engine, and the crossbar controller is a general peripheral for resistive arrays. Table~\ref{tab:overview} summarises the four blocks, their interfaces, and the headline implementation figures. The public repository of each block is pinned in the data-availability section to the commit described here.

\begin{table}[tbp]
\centering
\caption{Overview of the four blocks in the neuromorphic suite. Utilisation is the placed standard-cell utilisation on a single Tiny Tapeout tile, and tests are directed cocotb cases, all passing.}
\label{tab:overview}
\small
\begin{tabular}{@{}p{3.1cm}p{5.6cm}p{2.2cm}cc@{}}
\toprule
\rowcolor{headpastel}
\textbf{Block} & \textbf{Function} & \textbf{Interface} & \textbf{Util.} & \textbf{Tests}\\
\midrule
Ring-oscillator PVT sensor & Five selectable ring oscillators, frequency counter, jitter TRNG, and health monitor & SPI + parallel & 66\% & 29\\
\rowcolor{rowpastel}
Stochastic LIF neuron & LFSR-driven probabilistic integrate-and-fire with activation table and refractory period & SPI + parallel & 70\% & 18\\
STDP learning controller & Spike-timing plasticity with reward-modulated, eligibility-trace, and anti-Hebbian modes & SPI + spikes & 65\% & 22\\
\rowcolor{rowpastel}
Crossbar controller & Forming, set, reset, read and sweep with compliance limiting and half-select & SPI + analogue & 61\% & 30\\
\bottomrule
\end{tabular}
\end{table}

\subsection{Shared serial interface}\label{sec:spi}
\begin{figure}[tbp]
\centering
\includegraphics[width=0.92\textwidth]{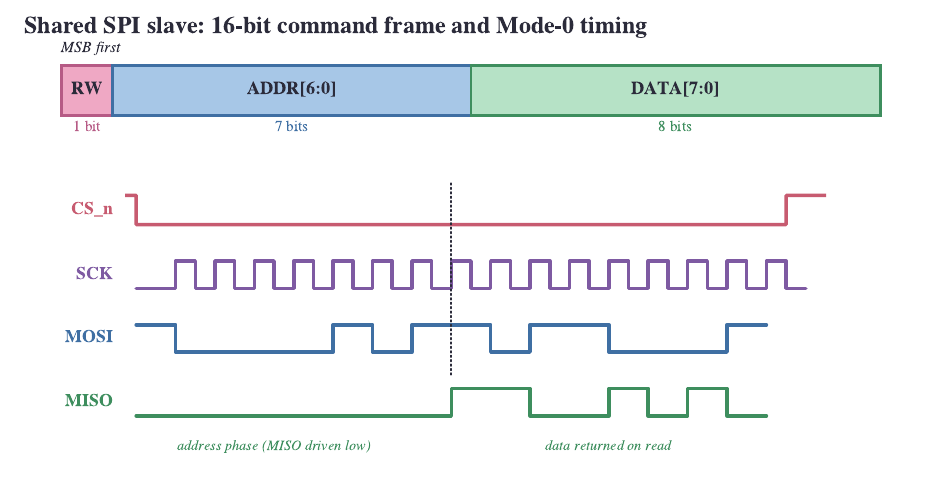}
\caption{Shared serial interface. The 16-bit command frame carries a read/write bit, a seven-bit address, and an eight-bit data field, transferred most significant bit first. The timing diagram shows a mode-0 transaction in which the address phase is presented on the host data line and read data is returned on the device data line during the second byte.}
\label{fig:spi}
\end{figure}

All four blocks share a single serial peripheral interface (SPI) slave module so that one host configures the whole suite; the sensor additionally exposes a direct parallel readout for bench measurement. The interface operates in mode 0, with the clock idle low and data sampled on the rising edge, and uses an active-low chip select. Each transaction is a 16-bit frame whose first bit selects read or write, the next seven bits carry the register address, and the final eight bits carry data, as shown in figure~\ref{fig:spi}. The fixed 16-register address space is sufficient for the control, status, and lookup-table contents of every block.

Because the serial clock is asynchronous to the 50\,MHz system clock, the slave passes the serial clock, chip select, and data inputs through multi-stage synchronisers before use, and detects clock edges from the synchronised samples. This removes metastability risk at the cost of a small fixed latency. The device data output is held in a high-impedance state whenever the chip select is inactive, implemented by gating the output-enable of the shared bidirectional pin with the chip-select signal. Bus isolation of this kind is necessary when several blocks occupy the same chip and share host wiring. The same slave module is instantiated unchanged in all four blocks, which keeps the host driver identical across them and reduces the verification surface to a single interface implementation.

\section{Process, voltage and temperature sensor}\label{sec:ro}
The sensor measures on-chip timing through five ring oscillators of 7, 11, 15, 21, and 31 inverting stages. Ring-oscillator frequency is a well established proxy for process and temperature variation, and compact oscillator-based sensors have been demonstrated across CMOS nodes, including the 0.13\,$\mu$m generation \cite{park2009a,woo2009a,wang2014a,rasaily2016cmos,rana2018cmos}. Each oscillator is built from SkyWater high-density standard cells, with a two-input NAND gate providing the enable and the start of the inversion chain and a series of inverters completing an odd number of inversions. Standard-cell instances are used rather than behavioural feedback because synthesis rejects combinational loops; the explicit cells with preservation attributes force the tool to keep the ring intact through optimisation and layout. The five lengths give five nominal frequencies from the same supply and temperature, which provides several reference points for characterising the operating corner when the counts are compared.

The selected oscillator output crosses into the system clock domain through a three-stage synchroniser, after which a 16-bit counter accumulates rising edges during a measurement window. The sensor supports two control modes, selected by a mode input (\texttt{ui\_in[6]}). In parallel-control mode the window is set externally: a count-enable input (\texttt{ui\_in[3]}) is held high for the gate interval, and a rising edge on a separate clear input (\texttt{ui\_in[4]}) latches the current count into a holding register and resets the counter; because the detector responds only to the rising edge, holding the clear line high does not continuously reset the count. In serial-control mode the host instead programs a 16-bit gate time (registers \texttt{0x03} and \texttt{0x04}) and starts a hardware auto-gate timer through the control register, which counts for the programmed interval, latches the result, and raises a measurement-done flag. The 16-bit result is read out one byte at a time under control of a byte-select input (\texttt{ui\_in[5]}), and an overflow flag indicates saturation of the counter. The raw and synchronised oscillator signals are also brought out so that frequency can be observed directly on a bench instrument.

Each oscillator can be disabled individually through the per-oscillator enable register, which power-gates the unused rings and allows the current of a single ring to be isolated for measurement. Selective gating supports characterisation across conditions: comparing the frequency codes of the five lengths, and of the same length across conditions, yields a compact signature of the operating corner. Resolving the individual process, voltage, and temperature contributions from these codes requires per-corner calibration, which is part of the planned silicon measurement. Figure~\ref{fig:pvt} shows the modelled frequency response of three representative oscillators against supply voltage and temperature using an alpha-power delay model, which illustrates the monotonic trends that the counter resolves.

\subsection{On-chip entropy}
A jitter-based true-random-number generator (TRNG) reuses the same oscillator bank. The phase jitter between two selected rings is sampled and combined through an exclusive-or network to extract random bits, a standard construction for ring-oscillator entropy sources \cite{sunar2007}. A control register enables the generator and a data register returns the sampled bits, so the host that reads the frequency counter can also collect entropy for stochastic computation or for seeding the neuron's shift register. Reusing one oscillator structure for both timing measurement and entropy keeps the added area small.

\subsection{Health and aging monitor}
A health monitor compares the measured frequency against programmable lower and upper bounds and raises an alert flag when the count falls outside the window or an oscillator stalls. Ring-oscillator frequency shift is an established proxy for circuit aging and degradation \cite{kim2008odometer}, so the same count that characterises the operating corner also reports drift over the life of the part. The bounds and the status flag are register-mapped, which lets a host poll the health of the die without external instrumentation.

\subsection{Differential and prescaled readout}
A differential mode selects a pair of rings and measures their beat frequency, which suppresses common-mode supply and temperature dependence and isolates a targeted difference. A configurable prescaler divides the selected oscillator by eight, sixteen, thirty-two, or sixty-four before the counter, trading resolution for range so that the fastest rings do not overflow the measurement window. The pair selection and the prescaler ratio are both set through the register file.

\subsection{Interface}
The sensor instantiates the same serial register slave as the other three blocks, so the oscillator selection, the entropy and health controls, and the count read-out are reached through the shared host protocol, while the raw count and oscillator signals remain available on parallel pins for direct bench measurement.

The sensor places at 66\% utilisation with a post-synthesis standard-cell area of about 9{,}950\,$\mu$m$^2$ and is documented with twenty-nine directed tests. The block is available at \url{https://github.com/santhoshs93/tt_um_santhosh_ring_osc}.

\begin{figure}[tbp]
\centering
\includegraphics[width=0.86\textwidth]{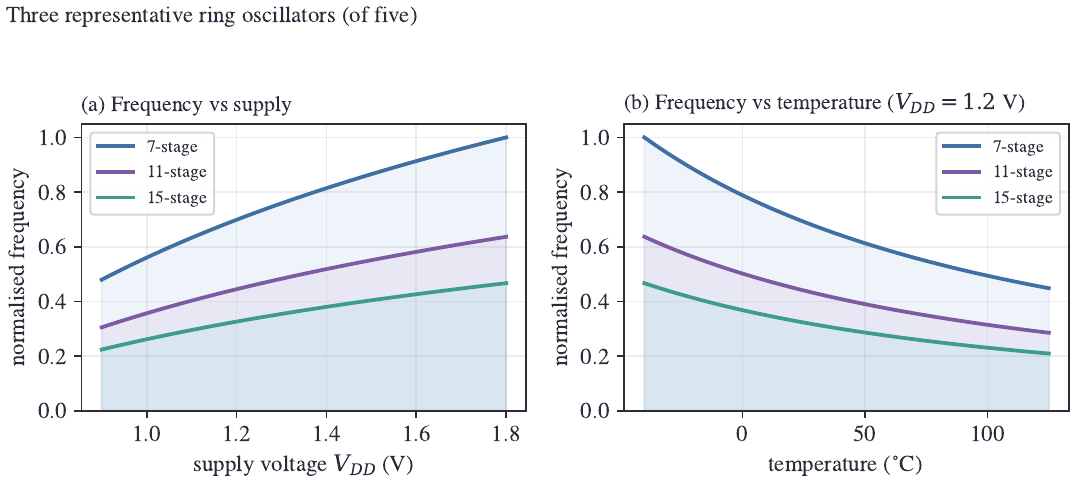}
\caption{Modelled response of three representative on-chip ring oscillators (7, 11, and 15 stages). (a) Normalised frequency, referenced to the seven-stage ring at 1.8\,V, against supply voltage at room temperature. (b) Normalised frequency, referenced to the seven-stage ring at the cold corner, against temperature at a 1.2\,V supply. The trends follow an alpha-power stage-delay model and indicate the monotonic dependence that the frequency counter is designed to resolve, once fabricated, for process, voltage and temperature characterisation.}
\label{fig:pvt}
\end{figure}

\section{Stochastic leaky integrate-and-fire neuron}\label{sec:neuron}
\begin{figure}[tbp]
\centering
\includegraphics[width=0.90\textwidth]{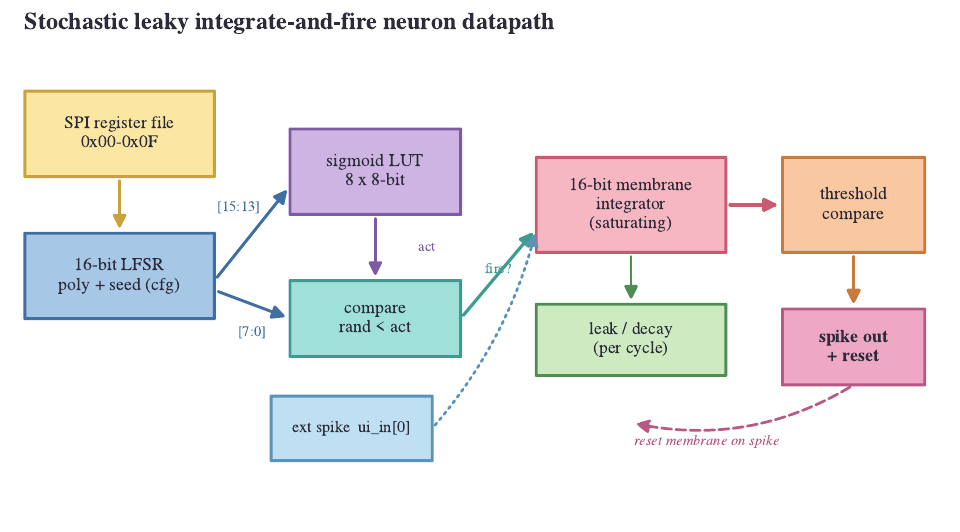}
\caption{Datapath of the stochastic leaky integrate-and-fire neuron. The configurable linear-feedback shift register supplies an address to the activation table and a comparison value. A successful comparison or an external spike drives the saturating membrane integrator, which leaks each idle cycle and emits a spike with reset when the upper byte of the membrane reaches the programmed threshold.}
\label{fig:neuron}
\end{figure}

The neuron combines a configurable pseudorandom source, a programmable activation lookup table (LUT), and a saturating leaky integrator. Its datapath is shown in figure~\ref{fig:neuron}. All parameters that shape its behaviour, namely the feedback polynomial, the seed, the activation table, the firing threshold, and the leak rate, are written through the serial register file, so the neuron is retuned in the field without changing the netlist.

\subsection{Pseudorandom source}
Randomness is generated by a 16-bit Fibonacci linear-feedback shift register. At each cycle the register shifts left and the new least significant bit is the parity of the bitwise product of the register state and the feedback polynomial,
\begin{equation}
b_{\mathrm{fb}} = \bigoplus_{i=0}^{15} \big( s_i \wedge p_i \big), \qquad s \leftarrow \big( (s \ll 1) \;\vert\; b_{\mathrm{fb}} \big)\, ,
\end{equation}
where $s$ is the register state and $p$ is the programmable polynomial. The polynomial value sets the sequence period: a maximal-length polynomial yields the full period of 65535 states, which is the configuration used for the characterisation in figure~\ref{fig:neurontransfer}, and an application that needs a long sequence should program such a polynomial. A guard forces the state back to a known non-zero value if it ever reaches all zeros, which removes the lock-up condition that would otherwise stall a Fibonacci register. Making the polynomial and seed writable lets the host trade sequence length against correlation properties, or load a known seed for repeatable behaviour.

\subsection{Activation and integration}
The three most significant bits of the register select one of eight entries in the activation table, and the eight least significant bits form a comparison value; the intermediate bits, $s_{12:8}$, do not participate in the comparison. The neuron registers a stochastic event when the comparison value is below the selected activation level,
\begin{equation}
\mathrm{fire}_{\mathrm{st}} = \big[\, s_{7:0} < \mathrm{LUT}[\,s_{15:13}\,] \,\big]\, ,
\end{equation}
so the firing probability of the stochastic path is set by the table contents. The default table approximates a sigmoid, which biases the neuron toward graded probabilistic activation, but any monotonic or non-monotonic shape can be programmed. The membrane is a 16-bit accumulator updated by
\begin{equation}
V[n+1] =
\begin{cases}
\min\big(V_{\max},\, V[n] + I\big), & \mathrm{fire}_{\mathrm{st}} \;\vee\; \mathrm{spike}_{\mathrm{ext}}\\[2pt]
\max\big(0,\, V[n] - \lambda\big), & \text{otherwise}
\end{cases}
\end{equation}
where $I$ is the input current, $\lambda$ is the programmable leak, and $V_{\max}$ is full scale. Saturating arithmetic is used in both directions so the membrane never wraps, which would otherwise produce spurious spikes. The input current is either the activation level in free-running mode or a four-bit parallel weight, zero-extended to the integrator width, in host-driven mode, selected by a mode input. A spike is emitted, and the membrane is reset to zero, when the upper byte of the membrane reaches the threshold register,
\begin{equation}
\mathrm{spike} = \big[\, V_{15:8} \geq \theta \,\big]\, .
\end{equation}
Comparing only the upper byte keeps the threshold path narrow and fixes the firing resolution at 256 membrane levels.

\subsection{Refractory period}
After a spike the neuron enters a refractory state for a programmable interval of up to seven cycles, during which further firing is suppressed and the membrane holds its value. The refractory length is set in the control register and a status bit reports when the neuron is refractory. A refractory period is a standard feature of integrate-and-fire models and bounds the maximum firing rate, which is useful when the stochastic path would otherwise produce closely spaced spikes. A digital leaky integrate-and-fire neuron with a pseudorandom stochastic path and a refractory period follows the established digital-neuron model used in large neurosynaptic cores \cite{cassidy2013}.

\subsection{Simulated dynamics}
\begin{figure}[tbp]
\centering
\includegraphics[width=0.74\textwidth]{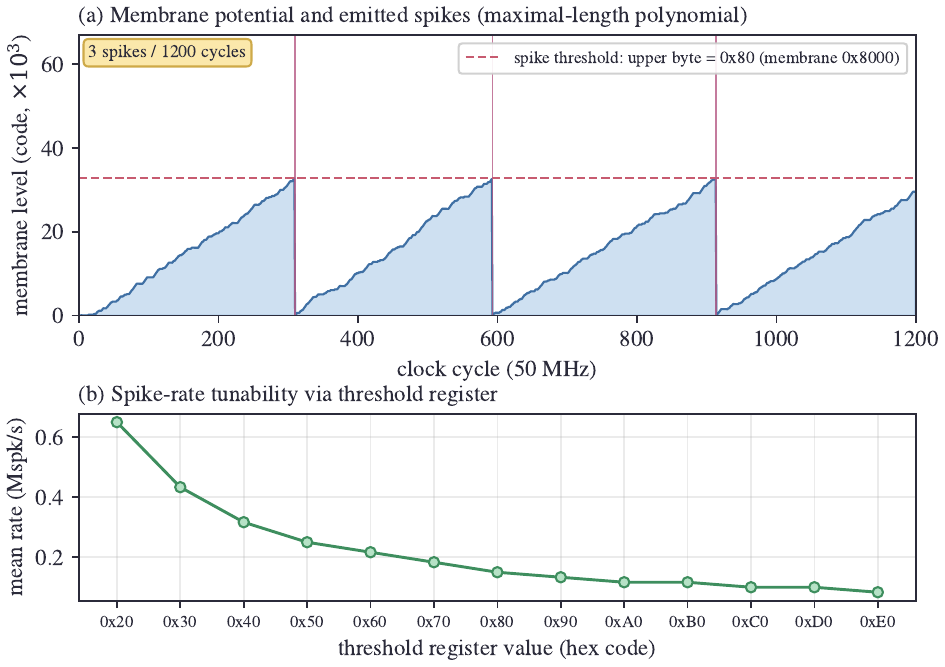}
\caption{Simulated neuron behaviour from the register-transfer model. (a) Membrane potential and emitted spikes using a maximal-length polynomial, showing accumulation toward the threshold followed by reset. (b) Mean firing rate against the threshold register value, demonstrating monotonic rate control over the configured range.}
\label{fig:neurondyn}
\end{figure}

\begin{figure}[tbp]
\centering
\includegraphics[width=0.92\textwidth]{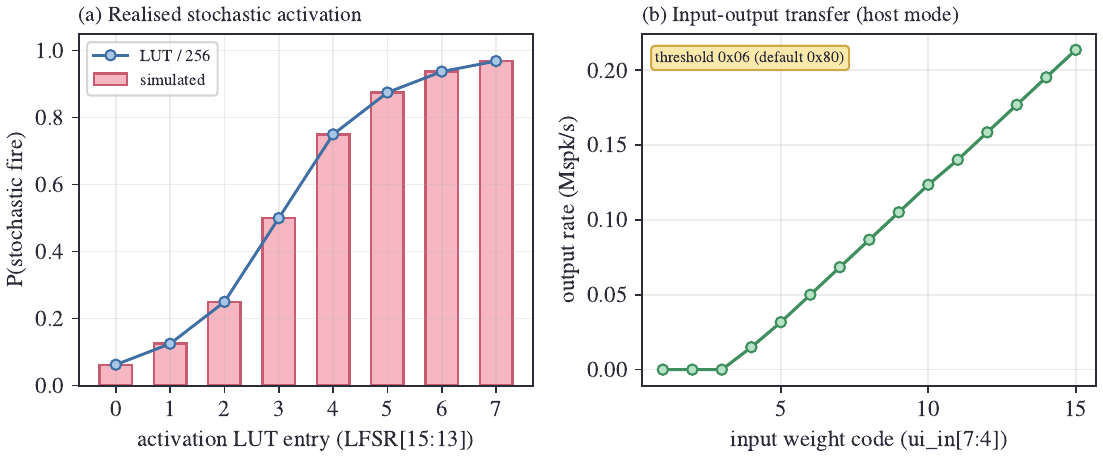}
\caption{Stochastic transfer behaviour of the neuron from the register-transfer model, using a maximal-length polynomial (period 65535). (a) The per-state stochastic-fire probability matches the programmed activation table divided by 256, a by-construction check that the pseudorandom source is uniform across states. (b) The mean output spike rate increases monotonically with the input weight code in host-driven mode at a reduced threshold.}
\label{fig:neurontransfer}
\end{figure}

Figure~\ref{fig:neurondyn} shows the behaviour of the block obtained by simulating the RTL model with the default configuration. The membrane potential ramps as stochastic events accumulate, crosses the threshold, and resets, giving the regular spike train in panel (a). Panel (b) sweeps the threshold register and reports the mean firing rate, which falls monotonically as the threshold rises. The threshold register therefore provides direct, monotonic control of excitability, and the leak and activation table provide independent control of integration and of the stochastic firing probability. Figure~\ref{fig:neurontransfer} characterises this computation directly: with a maximal-length polynomial the per-state firing probability matches the programmed activation divided by 256, and the output spike rate is a monotonic function of the input weight. At 70\% tile utilisation the neuron is the densest of the four blocks, a consequence of its 16-bit datapath and table storage; its post-synthesis standard-cell area is about 10{,}590\,$\mu$m$^2$, and eighteen directed tests cover it. The block is available at \url{https://github.com/santhoshs93/tt_um_santhosh_stoch_neuron}.

\section{On-chip STDP learning controller}\label{sec:stdp}
\begin{figure}[tbp]
\centering
\includegraphics[width=0.90\textwidth]{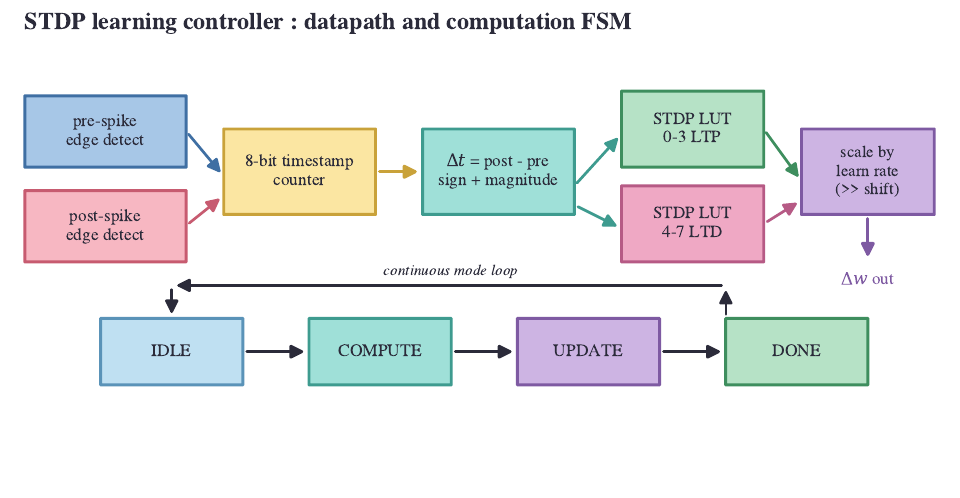}
\caption{STDP learning controller. Pre- and post-synaptic timestamps are captured against an eight-bit counter, their signed difference indexes a programmable plasticity table split into potentiation and depression halves, and the result is scaled by a learning rate. A four-state machine sequences the computation and supports single-shot and continuous operation.}
\label{fig:stdp}
\end{figure}

The learning controller turns spike timing into weight changes using a programmable plasticity curve. Its datapath and control sequence are shown in figure~\ref{fig:stdp}. An external timestamp clock drives an eight-bit counter, and two-stage synchronisers detect the rising edges of the pre- and post-synaptic spike inputs. When a spike edge is detected, the current timestamp is captured into the corresponding register and a valid flag is set.

When both timestamps are valid and learning is enabled, a four-state finite-state machine (FSM), with states for idle, compute, update, and done, performs the calculation. It first forms the signed time difference
\begin{equation}
\Delta t = t_{\mathrm{post}} - t_{\mathrm{pre}}\, .
\end{equation}
The sign selects potentiation or depression and the magnitude indexes the plasticity table, whose first four entries hold the long-term potentiation (LTP) curve and whose last four hold the long-term depression (LTD) curve. Because a zero time difference is classified as potentiation, the zero-magnitude depression entry is not exercised, and the reachable depression entries are those for differences of one to three. The magnitude is taken from the table and scaled by a programmable learning rate implemented as a right shift,
\begin{equation}
\Delta w = \mathrm{LUT}[k] \,\gg\, (3 - r), \qquad
k =
\begin{cases}
|\Delta t|, & \Delta t \geq 0 \;\;(\text{potentiation})\\
4 + |\Delta t|, & \Delta t < 0 \;\;(\text{depression})
\end{cases}
\end{equation}
where the two-bit rate field $r$ maps to division by eight, four, or two, or to the full value. Only the low two bits of the learning-rate register and of the magnitude are used, so the rate has four settings and the curve resolves four magnitude bins; the learning window is therefore intended to span time differences of at most three, and because the magnitude index uses the low two bits, a window programmed above three would alias larger differences onto the same entries. The table lookup and the shift occupy successive states of the computation machine, so the expression above describes the settled update. A configurable time window suppresses updates for spike pairs that fall outside the learning interval, and an overflow flag reports when a table entry is at full scale so that the host can detect saturation of the synaptic weight. The controller supports a single-shot mode, in which it holds the result until reset, and a continuous mode, in which it returns to idle and processes the next spike pair.

\subsection{Three-factor extensions}
\begin{figure}[tbp]
\centering
\includegraphics[width=0.92\textwidth]{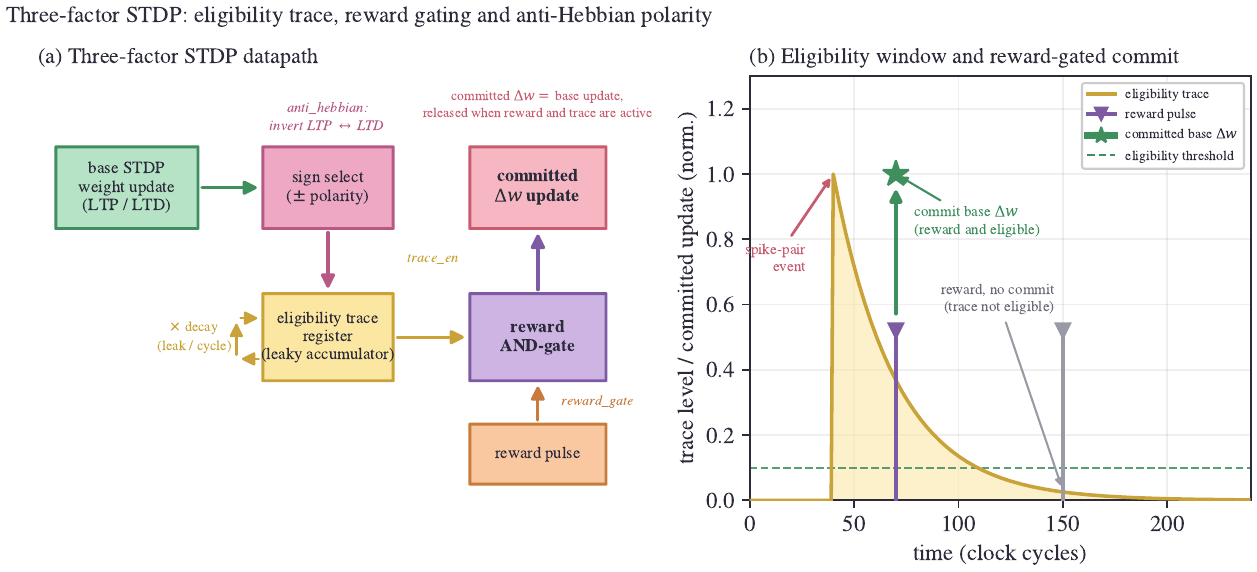}
\caption{Three-factor extensions of the learning controller. (a) The base spike-timing update passes through an optional sign inversion for the anti-Hebbian mode and is then released by a reward gate that is enabled only while the eligibility trace is non-zero, so the committed change is the base update gated by reward and recent activity rather than a function of the decayed trace level. (b) The eligibility trace rises on a spike pair and then decays; a reward pulse that arrives while the trace is still above the eligibility threshold releases the base update, whereas a reward after the trace has decayed commits nothing.}
\label{fig:stdpthree}
\end{figure}

Beyond pair-based potentiation and depression the controller implements three modes that are widely used in reward-driven and competitive learning. Each is established in the literature and is provided here as a configurable digital option rather than as a new rule. In reward-modulated mode the signed weight update is gated by an external reward input, so a synapse changes only when a global reward coincides with the eligible spike pair; this is the standard reward-modulated spike-timing-dependent plasticity (R-STDP) rule \cite{izhikevich2007,florian2007}. An eligibility trace, implemented as a leaky counter that decays at a programmable rate, holds the recent coincidence so that a reward arriving after a short delay still credits the correct pair, the mechanism that links spike timing to delayed reward in three-factor learning rules \cite{legenstein2008,fremaux2016}. An anti-Hebbian mode inverts the sign of the update, exchanging potentiation and depression, which supports decorrelation and competition. The reward gate, the trace enable and decay rate, and the anti-Hebbian select are register bits, and figure~\ref{fig:stdpthree} shows how they combine with the base update. A compact digital implementation of reward-modulated plasticity with an eligibility trace has been reported previously \cite{quintana2022}; the controller here follows that line and exposes the behaviour through the shared register interface.

\begin{figure}[tbp]
\centering
\includegraphics[width=0.86\textwidth]{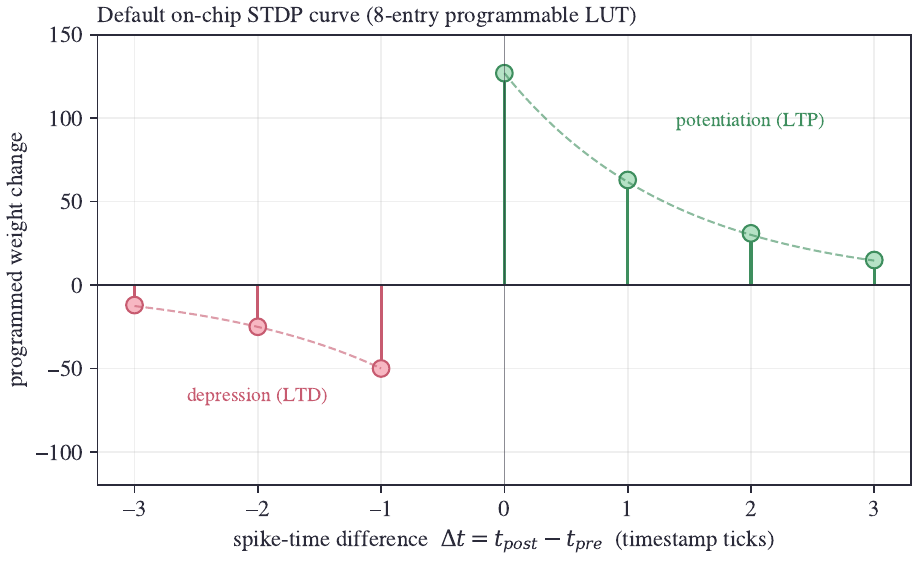}
\caption{Default on-chip plasticity curve held in the eight-entry table. Potentiation entries apply for a positive spike-time difference and depression entries for a negative difference. The values implement an asymmetric characteristic; the dashed curves are guides to the eye for the default entries, which are fully programmable.}
\label{fig:stdpcurve}
\end{figure}

Storing the plasticity curve in a writable table separates the learning rule from the synaptic medium, so the same controller drives any device technology by reprogramming the curve. Figure~\ref{fig:stdpcurve} shows the default curve, which gives a strong potentiation peak for near-coincident pre-before-post pairs and a weaker depression lobe for post-before-pre pairs, the asymmetric shape commonly used in unsupervised feature learning. The block places at 65\% utilisation with a post-synthesis standard-cell area of about 9{,}850\,$\mu$m$^2$ and is covered by twenty-two directed tests. It is available at \url{https://github.com/santhoshs93/tt_um_santhosh_stdp_ctrl}.

\section{Memristive crossbar programming controller}\label{sec:xbar}
\begin{figure}[tbp]
\centering
\includegraphics[width=0.90\textwidth]{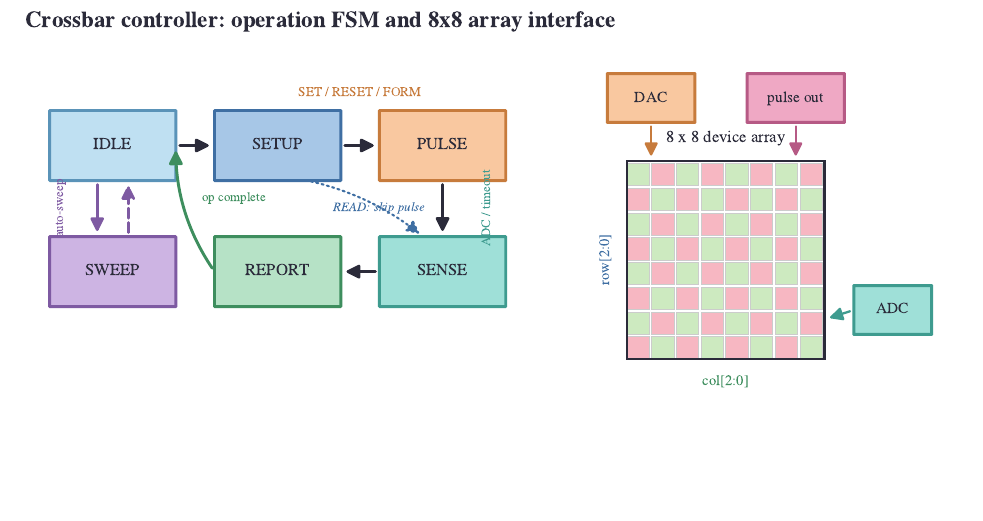}
\caption{Crossbar controller operation. A seven-state machine sequences forming, set, reset, read, and an automated sweep over an eight-by-eight device array, with an inter-pulse gap state for pulse trains. Programming pulses of configurable width are applied through a digital-to-analogue code, and an external analogue-to-digital converter returns the sensed value with a ready handshake or a timeout.}
\label{fig:xbar}
\end{figure}

The crossbar controller is a device-agnostic peripheral for resistive arrays. It addresses an eight-by-eight array with three-bit row and column fields and supports four operations, namely read, set, reset, and forming, selected through a mode register. A seven-state machine, with states for idle, setup, pulse, gap, sense, report, and sweep, sequences each operation, as shown in figure~\ref{fig:xbar}. On a start command the machine asserts the row and column enables, then either applies a programming pulse for set, reset, and forming, or proceeds directly to sensing for a read.

\begin{figure}[tbp]
\centering
\includegraphics[width=0.92\textwidth]{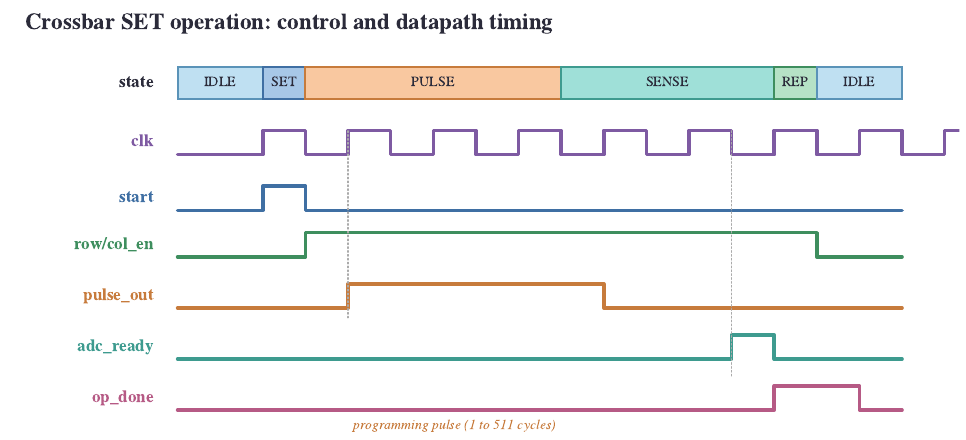}
\caption{Control and datapath timing of a crossbar SET operation. After the start command the controller asserts the row and column enables, drives the programming pulse for the configured width, waits in the sense phase for the converter ready handshake, and reports completion before returning to idle. A read operation follows the same sequence but skips the pulse phase.}
\label{fig:xbartiming}
\end{figure}

The programming pulse has a nine-bit configurable width, giving durations from a single cycle (20\,ns at the 50\,MHz clock) up to 511 cycles (about 10\,$\mu$s). The voltage level is set by an eight-bit digital-to-analogue converter (DAC) code; the two least significant bits drive dedicated output pins for an external converter, and the full code is available through the register file. During the sense phase the controller waits for an external analogue-to-digital converter (ADC) ready handshake and captures the four-bit reading, which it stores, nibble-replicated (the four bits are copied into both halves of the byte), in the eight-bit read register. A 256-cycle timeout protects against a missing converter response and raises an error flag so that the host is not left waiting indefinitely. Figure~\ref{fig:xbartiming} shows the control timing of a SET operation through these phases.

For current-voltage characterisation the controller provides an automated sweep mode that steps the voltage code from a start value to an end value by a programmable increment, applying a pulse at each step. A guard terminates the sweep when the increment is zero, which prevents an unintended infinite loop, and an abort control returns the machine to idle and clears the outputs at any time. Exposing pulse width, voltage code, and sweep bounds as register fields keeps the controller independent of the device under test, so resistive, ferroelectric, and spintronic arrays with different programming requirements are all driven by reconfiguration rather than redesign. \subsection{Compliance limiting}
During a set or forming pulse the controller monitors the sensed current and aborts the pulse when it exceeds a programmable compliance threshold, which protects the device from the runaway current that would drive a filament past its target state. Current compliance is the standard means of controlling resistive switching \cite{alibart2012}, and exposing the threshold as a register lets the host match it to the device under test. A status bit records that compliance was reached so the host can respond.

\subsection{Half-select biasing}
To address one device without disturbing the rest of the array, the controller drives the unselected lines to an intermediate bias set by a digital-to-analogue code, the half-select scheme that limits leakage through partially selected paths. Sneak-path current is a well-known limit in resistive crossbars and bias schemes of this kind are a common mitigation \cite{cassuto2016}. The bias code is register-mapped and read back for calibration.

\subsection{Pulse-train programming}
For devices that need repeated pulses the controller issues a programmable train, with a repeat count and an inter-pulse gap, and counts the pulses actually delivered in a read-only register. Incremental pulse-and-verify programming of this kind is standard practice for resistive memory \cite{alibart2012}. The repeat count, the gap, and the per-pulse width are register fields, and the abort control stops the train at any point.

At 61\% utilisation and a post-synthesis standard-cell area of about 9{,}250\,$\mu$m$^2$, the crossbar controller carries the largest test count in the suite, thirty directed cases, reflecting its several operating modes. It is available at \url{https://github.com/santhoshs93/tt_um_santhosh_xbar_ctrl}.

\section{Programming model and register maps}\label{sec:prog}
\begin{figure}[tbp]
\centering
\includegraphics[width=0.84\textwidth]{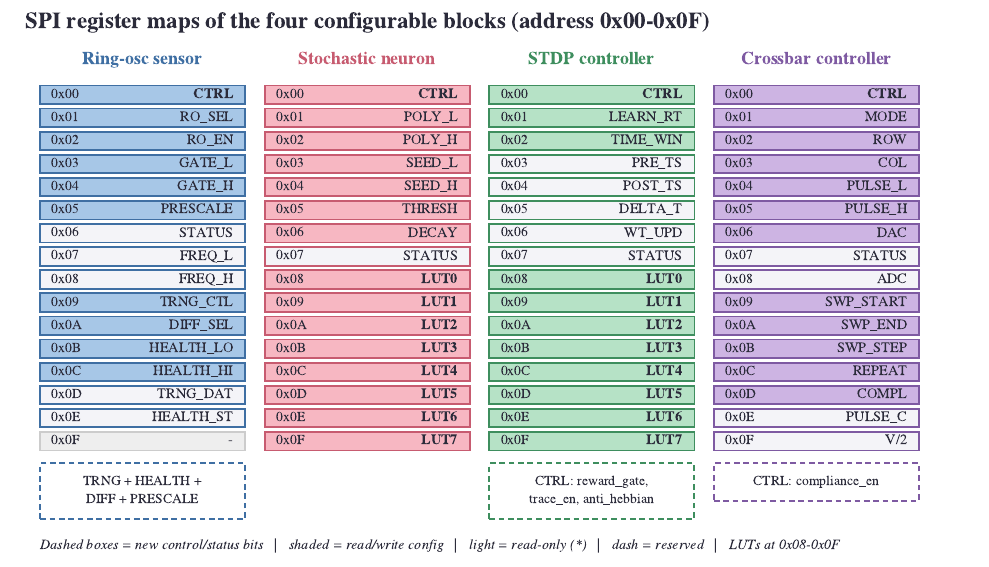}
\caption{Serial register maps of the four blocks. A common layout places a control register at address zero and a status register near the top of the lower half (address seven for the three programmable blocks, address six for the sensor), with read-only result registers and the eight-entry lookup tables occupying the upper half of the address space where present.}
\label{fig:regmap}
\end{figure}

Each block presents a 16-entry register file at a fixed address layout, shown in figure~\ref{fig:regmap}. The first register is always a control register whose bits start, reset, or select the operating mode of the block. A status register reports completion, error, and internal flags. The remaining registers hold block-specific configuration, and the upper half of the address space, from address eight to address fifteen, holds the eight-entry lookup table where the block defines one. This regularity means that a host driver discovers and configures every block through the same access pattern, and that the activation table of the neuron, the plasticity table of the learning controller, and the operating registers of the crossbar controller are all reached by the same sequence of serial writes. Table~\ref{tab:bitfields} lists the control, status, and mode bit fields used to operate the four blocks; the full per-pin assignment for each block is provided with the source.

\begin{table}[tbp]
\centering
\caption{Control, status, and mode register bit fields for the four blocks.}
\label{tab:bitfields}
\small
\begin{tabular}{@{}llp{8.0cm}@{}}
\toprule
\rowcolor{headpastel}
\textbf{Block} & \textbf{Register} & \textbf{Bit fields}\\
\midrule
Sensor & CTRL (0x00) & [0] auto-gate start, [1] clear measurement; RO select at 0x01, prescaler at 0x05\\
\rowcolor{rowpastel}
Sensor & TRNG/health (0x09) & [0] TRNG enable, [1] health enable, [2] differential mode; bounds at 0x0B and 0x0C, status at 0x0E\\
Neuron & CTRL (0x00) & [0] enable, [1] accumulator reset, [2] free-run, [7:5] refractory period\\
\rowcolor{rowpastel}
Neuron & STATUS (0x07) & [0] spike, [1] accumulator overflow, [2] spike latched, [3] membrane MSB, [6] refractory\\
STDP & CTRL (0x00) & [0] enable, [1] reset, [2] single-shot, [3] anti-Hebbian, [4] reward gate, [5] eligibility trace\\
\rowcolor{rowpastel}
STDP & STATUS (0x07) & [0] update ready, [1] potentiation, [2] depression, [3] weight overflow, [4] reward, [5] trace non-zero\\
Crossbar & CTRL (0x00) & [0] start, [1] abort, [2] auto-sweep, [3] compliance enable\\
\rowcolor{rowpastel}
Crossbar & MODE (0x01) & [1:0] operation: 00 read, 01 set, 10 reset, 11 forming\\
Crossbar & STATUS (0x07) & [0] busy, [1] operation done, [2] operation error, [3] sense input, [7] compliance hit\\
\bottomrule
\end{tabular}
\end{table}

A typical configuration sequence writes the control register last, after the operating parameters and any table contents are in place, so that the block begins from a defined state. For the neuron the host writes the polynomial, seed, threshold, and leak, optionally programs the activation table, and then enables the block. For the learning controller the host sets the learning rate and time window, optionally programs the plasticity table, and enables learning before supplying the timestamp clock and the spike inputs. For the crossbar controller the host writes the mode, the row and column addresses, the pulse width, and the voltage code, then sets the start bit, after which the sequence of pulse and sense phases runs to completion and reports through the status register.

\section{Functional verification}\label{sec:verif}
Verification used the cocotb Python-based hardware verification framework (cocotb 2.0.1) with the Icarus Verilog simulator and a total of 99 directed test cases across the four blocks, all passing, distributed as shown in table~\ref{tab:tests}. The tests exercise reset behaviour, register read and write paths, the stochastic and integration logic of the neuron, the timestamp capture and plasticity computation of the learning controller, every operating mode and the sweep guard of the crossbar controller, and the edge-detected control and counter behaviour of the sensor. Directed cases were chosen to cover the corner conditions that the design decisions target, for example the all-zero lock-up guard of the shift register, the saturation limits of the membrane, the time-window boundary of the learning rule, and the zero-increment sweep guard.

\begin{table}[tbp]
\centering
\caption{Directed cocotb test cases per block, run at register-transfer level and, where applicable, at gate level on the implemented netlist.}
\label{tab:tests}
\small
\begin{tabular}{@{}p{6.0cm}cc@{}}
\toprule
\rowcolor{headpastel}
\textbf{Block} & \textbf{Tests} & \textbf{Gate-level}\\
\midrule
Ring-oscillator PVT sensor & 29 & see text\\
\rowcolor{rowpastel}
Stochastic LIF neuron & 18 & yes\\
STDP learning controller & 22 & yes\\
\rowcolor{rowpastel}
Crossbar controller & 30 & yes\\
\midrule
\textbf{Total} & \textbf{99} & \\
\bottomrule
\end{tabular}
\end{table}

The same test benches were re-run at gate level on the implemented netlists of the three synchronous blocks, which confirms that the behaviour observed at register-transfer level survives synthesis and layout. Gate-level simulation of the oscillator block is treated separately because the ring oscillators are intentional combinational loops; their oscillation is the physical behaviour under measurement rather than a timing fault, and the register-transfer model substitutes clock dividers for the rings during logic simulation so that the counter and readout logic can still be checked deterministically. Each block carries a continuous-integration workflow that runs the test suite, builds the documentation, generates the layout, and runs the design-rule and layout-versus-schematic checks, so that any change is re-verified automatically.

\section{Physical implementation and results}\label{sec:impl}
The four blocks were implemented on the SkyWater 130\,nm process using an open-source digital integrated-circuit design flow based on OpenLane and OpenROAD, covering standard-cell synthesis, placement, clock-tree synthesis, and sign-off, with the parameters summarised in table~\ref{tab:impl}. Each block is placed on a single tile with an absolute die size, a target placement density of 60\% at the pinned commits, and a single power-distribution layer appropriate to the tile. This density target is the value supplied to the placer and is distinct from the achieved per-block core utilisation reported below. Clock-tree synthesis is enabled, and the static timing analysis in the flow was constrained to the 50\,MHz system clock with hold-slack margins applied at placement and routing. All four blocks complete the flow meeting the 50\,MHz timing constraint reported by the static timing analysis in the flow, pass the design-rule and layout-versus-schematic checks, and produce a manufacturable layout. The results in this section come from the implementation flow and from functional simulation; no fabricated silicon is reported, and electrical characterisation is outlined in section~\ref{sec:disc}.

\begin{table}[tbp]
\centering
\caption{Physical implementation parameters and outcomes common to the four blocks.}
\label{tab:impl}
\small
\begin{tabular}{@{}p{5.0cm}p{8.2cm}@{}}
\toprule
\rowcolor{headpastel}
\textbf{Parameter} & \textbf{Value}\\
\midrule
Process & SkyWater 130\,nm open process design kit\\
\rowcolor{rowpastel}
Tile footprint & single Tiny Tapeout 1x1 tile, about 161\,$\mu$m by 112\,$\mu$m\\
Target clock & 50\,MHz (20\,ns period)\\
\rowcolor{rowpastel}
Placement density target & 60\%\\
Flow & OpenLane 2 and OpenROAD standard-cell flow (sky130A)\\
\rowcolor{rowpastel}
Sign-off & design-rule and layout-versus-schematic clean\\
Verification & 99 cocotb cases (cocotb 2.0.1, Icarus Verilog), register-transfer and gate level\\
\bottomrule
\end{tabular}
\end{table}

\begin{figure}[tbp]
\centering
\includegraphics[width=0.86\textwidth]{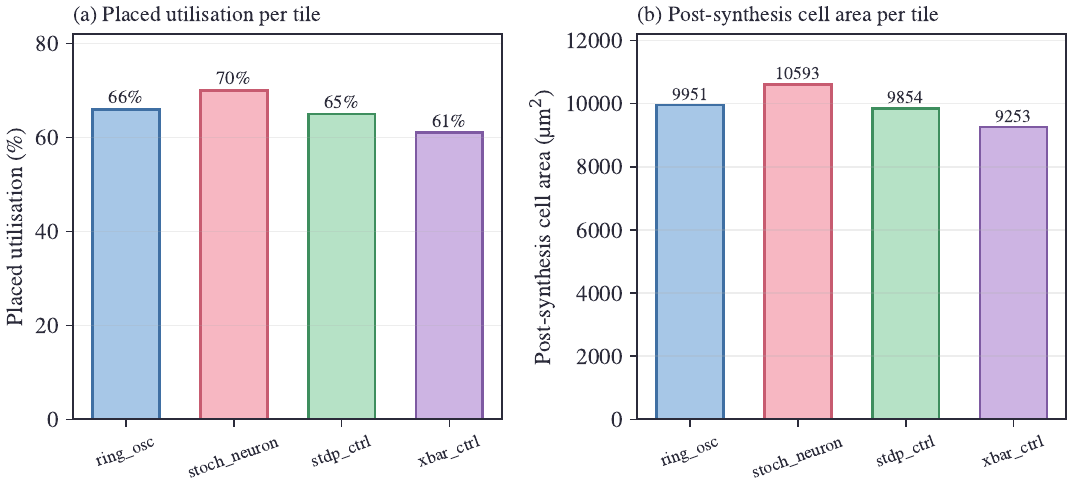}
\caption{Implementation summary from the open standard-cell flow. (a) Placed standard-cell utilisation per tile. (b) Post-synthesis standard-cell area per tile in square micrometres.}
\label{fig:util}
\end{figure}

\begin{figure}[tbp]
\centering
\includegraphics[width=0.82\textwidth]{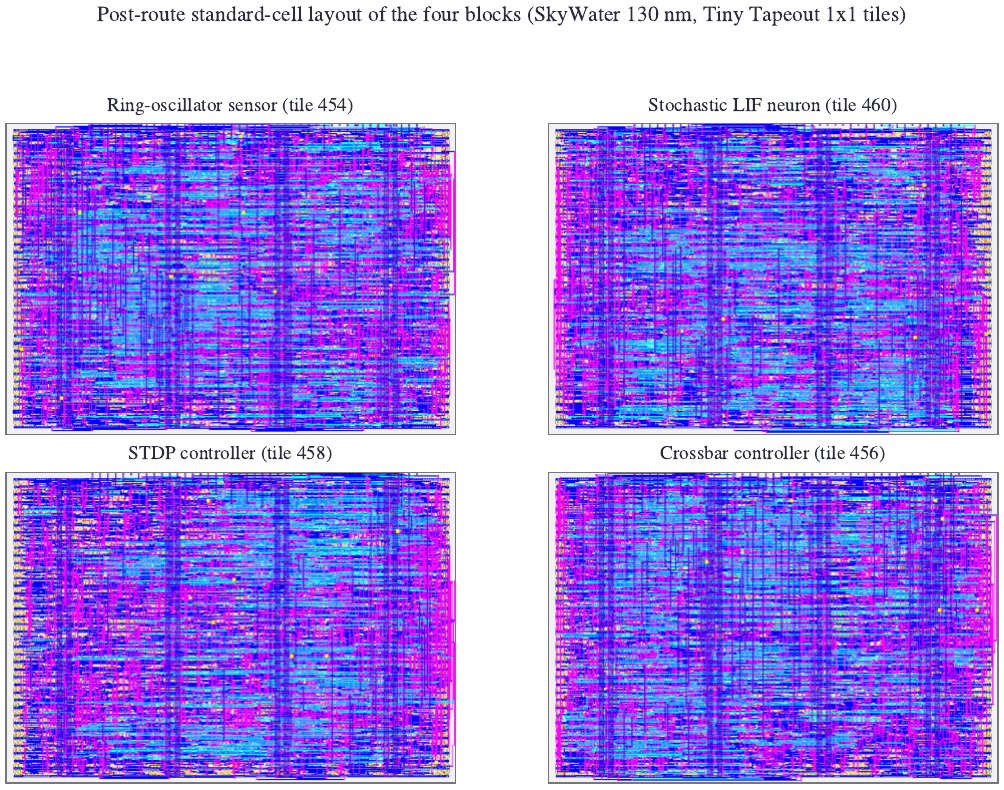}
\caption{Post-route standard-cell layout of the four blocks on the SkyWater 130\,nm process. Each block is hardened as a single Tiny Tapeout 1x1 tile (about 161 by 112\,$\mu$m) and rendered from the routed GDSII produced by the open flow on the ttsky26a shuttle. These are layout renders, not measured silicon.}
\label{fig:floorplan}
\end{figure}

Figure~\ref{fig:util} reports the placed utilisation and the post-synthesis cell area for each block, table~\ref{tab:ppa} gives the full per-block implementation summary, and figure~\ref{fig:floorplan} shows the post-route layout of the four blocks rendered from the routed GDSII. Placed utilisation ranges from 61\% for the crossbar controller to 70\% for the neuron, with the sensor at 66\% and the learning controller at 65\%, and the corresponding post-synthesis standard-cell areas span 9{,}253 to 10{,}593\,$\mu$m$^2$ (table~\ref{tab:ppa}). The ordering follows the datapath width and the amount of stored state: the neuron carries a 16-bit shift register, a 16-bit membrane, and an activation table; the learning controller carries timestamp logic, a state machine, and a plasticity table; the crossbar controller carries a pulse counter, a sweep datapath, and a state machine; and the sensor carries only the oscillators, a synchroniser, and a counter. The headroom in the lower-utilisation tiles leaves room for additional channels or wider counters within the same footprint.

\begin{table}[tbp]
\centering
\caption{Per-block implementation summary on the SkyWater 130\,nm open process, reproduced in this work with Yosys, OpenROAD, and OpenSTA on the open \texttt{sky130\_fd\_sc\_hd} cell library. Cell count, flip-flop count, and standard-cell area are post-synthesis; placed utilisation, worst-case setup and hold slack, and power are from an open placement, clock-tree, and static-timing flow on a single Tiny Tapeout tile at 50\,MHz. Power is a pre-silicon estimate at a default switching activity and is not a measured value.}
\label{tab:ppa}
\small
\begin{tabular}{@{}lccccccc@{}}
\toprule
\rowcolor{headpastel}
\textbf{Block} & \textbf{Cells} & \textbf{FFs} & \textbf{Area} & \textbf{Util.} & \textbf{Setup} & \textbf{Hold} & \textbf{Power}\\
\rowcolor{headpastel}
 & & & ($\mu$m$^2$) & (\%) & (ns) & (ns) & ($\mu$W)\\
\midrule
Ring-oscillator sensor & 859 & 219 & 9{,}951 & 66 & +15.7 & +0.33 & 675\\
\rowcolor{rowpastel}
Stochastic LIF neuron & 917 & 208 & 10{,}593 & 70 & +14.3 & +0.35 & 701\\
STDP controller & 833 & 212 & 9{,}854 & 65 & +16.5 & +0.34 & 678\\
\rowcolor{rowpastel}
Crossbar controller & 840 & 198 & 9{,}253 & 61 & +15.9 & +0.33 & 644\\
\midrule
\textbf{Total} & \textbf{3{,}449} & \textbf{837} & \textbf{39{,}651} & n/a & n/a & n/a & \textbf{2{,}698}\\
\bottomrule
\end{tabular}
\end{table}

Table~\ref{tab:compare} positions the suite against representative neuromorphic hardware. The comparison is at the level of capability coverage rather than of a single block, since the cited systems differ in node and scale. Entries in table~\ref{tab:compare} reflect capabilities as described in each cited work; because the systems target different nodes and scales, the comparison is qualitative. Table~\ref{tab:open} positions the suite against open-source neuromorphic releases, which are individual cores, whereas the present work provides several primitives behind one register interface.

\begin{table}[tbp]
\centering
\caption{Qualitative positioning against representative neuromorphic hardware. The cited systems target different technology nodes and scales, so this feature-level comparison is qualitative rather than a measured benchmark.}
\label{tab:compare}
\small
\begin{tabular}{@{}p{4.4cm}cccc@{}}
\toprule
\rowcolor{headpastel}
\textbf{Aspect} & \textbf{Loihi}~\cite{davies2018loihi} & \textbf{Analogue neuron}~\cite{srivastava2016silico} & \textbf{STDP synapse}~\cite{mostafa2014a} & \textbf{This suite}\\
\midrule
Stochastic inference & partial & no & no & yes\\
\rowcolor{rowpastel}
On-chip plasticity & yes & no & yes & yes\\
Crossbar programming & no & no & no & yes\\
\rowcolor{rowpastel}
On-chip PVT sensing & no & no & no & yes\\
Shared register model & yes & no & no & yes\\
\rowcolor{rowpastel}
Open RTL and physical-design flow & no & no & no & yes\\
\bottomrule
\end{tabular}
\end{table}

\begin{table}[tbp]
\centering
\caption{Positioning against open-source neuromorphic hardware releases. Scope indicates whether the release is a single monolithic core or a set of separately addressable blocks. Node and capabilities are as reported by each cited work; the present work is the only entry that exposes heterogeneous primitives through one shared register interface.}
\label{tab:open}
\small
\begin{tabular}{@{}p{3.0cm}p{1.9cm}p{1.5cm}p{2.3cm}p{2.7cm}@{}}
\toprule
\rowcolor{headpastel}
\textbf{Release} & \textbf{Node} & \textbf{Open PDK} & \textbf{On-chip learning} & \textbf{Scope}\\
\midrule
OpenSpike \cite{modaresi2023} & sky130 130\,nm & yes & no & single SNN accelerator\\
\rowcolor{rowpastel}
ODIN \cite{frenkel2019} & 28\,nm & no & yes, SDSP & single 256-neuron core\\
ReckON \cite{frenkel2022} & 28\,nm & no & yes, e-prop & single recurrent SNN core\\
\rowcolor{rowpastel}
NeuroCoreX \cite{neurocorex2025} & FPGA & n/a & yes, STDP & single SNN emulator\\
This suite & sky130 130\,nm & yes & yes, STDP and R-STDP & four primitives, one interface\\
\bottomrule
\end{tabular}
\end{table}

The control logic also fixes the timing of each operation. Table~\ref{tab:timing} lists these operation latencies as deterministic cycle counts from the register-transfer and gate-level models, in clock cycles and in time at the 50\,MHz clock; they are design latencies, not measured silicon throughput.

\begin{table}[tbp]
\centering
\caption{Operation timing from the register-transfer and gate-level models, as deterministic cycle counts and as time at the 50\,MHz system clock (20\,ns period). These are design latencies of the control logic, not measured silicon throughput. The serial register interface is clocked independently of the system clock, so its access latency is given in serial-clock periods.}
\label{tab:timing}
\small
\begin{tabular}{@{}p{6.4cm}p{3.0cm}p{3.4cm}@{}}
\toprule
\rowcolor{headpastel}
\textbf{Operation} & \textbf{Cycles} & \textbf{Time at 50\,MHz}\\
\midrule
Serial register access (16-bit frame) & 16 serial-clock & 16 serial-clock periods\\
\rowcolor{rowpastel}
Neuron membrane update & 1 per cycle & 20\,ns\\
Neuron refractory hold & 0 to 7 & 0 to 140\,ns\\
\rowcolor{rowpastel}
STDP weight-update compute & 3 & 60\,ns\\
Crossbar programming pulse & 1 to 511 & 20\,ns to 10.2\,$\mu$s\\
\rowcolor{rowpastel}
Crossbar sense timeout & 256 & 5.12\,$\mu$s\\
Sensor gate (measurement) window & 1 to 65535 & 20\,ns to 1.31\,ms\\
\bottomrule
\end{tabular}
\end{table}

\FloatBarrier
\section{Discussion}\label{sec:disc}

\subsection{Design trade-offs}
Several decisions keep each block inside a single tile and a 50\,MHz budget. The lookup tables are limited to eight entries because wider tables overflow the tile, and the eight points are sufficient to define a useful activation or plasticity shape. The membrane threshold compares only the upper byte, which narrows the comparison path at the cost of coarser threshold resolution. Reads from the lookup tables index the storage array directly rather than a registered output, which avoids a stale-data hazard introduced by the non-blocking update of the register file. The oscillator clear is edge-detected so that a held control line does not continuously reset the counter. Each of these choices trades a small amount of generality for the area and timing closure required by the target.

\subsection{Limitations}
The results reported here are pre-silicon. Energy per operation, the absolute oscillation frequency of the rings, the metastability margin of the asynchronous interfaces under real timing, and the behaviour of an attached device array are not established by simulation and the implementation flow alone, and are deferred to electrical measurement. The area, utilisation, and timing reported here are from synthesis, placement, and clock-tree synthesis; routed sign-off with parasitic-extracted timing is not included, and the power figure is a pre-silicon estimate at a default switching activity rather than an activity-annotated or measured value. The four blocks are verified per tile rather than as a single co-simulated netlist, so the pipeline of figure~\ref{fig:system} describes an intended dataflow across interface-compatible blocks rather than a fabricated end-to-end system. The modelled oscillator response of figure~\ref{fig:pvt} is an analytical trend rather than a circuit simulation. Full static-timing reports, corner settings, and per-test coverage are produced by the continuous-integration flow rather than reproduced here.

\subsection{Toward silicon characterisation}
The results reported here are from functional verification, the implementation flow, and behavioural simulation of the register-transfer models. The next stage is electrical measurement of fabricated devices, for which the blocks already expose the necessary observability. The oscillator block brings out raw and synchronised oscillator signals and a 16-bit count for direct frequency measurement across supply and temperature, which yields the measured PVT response that figure~\ref{fig:pvt} models. The neuron exposes the membrane upper bits and spike output for firing-rate measurement against the threshold sweep of figure~\ref{fig:neurondyn}. The learning controller exposes the computed difference and weight update for verification of the programmed curve against figure~\ref{fig:stdpcurve}. Energy per operation and per spike, which depend on switching activity and the delivered supply, are the primary quantities to extract from measurement and are left to that stage rather than estimated here.

\subsection{Scaling and integration}
The blocks are written as standard-cell RTL with no full-custom analogue content except the oscillator cells, so they port to other nodes and to denser tiles by re-running the flow. The crossbar controller is the natural integration point for emerging memory: its register-defined pulse width, voltage code, and sweep parameters address the programming controls reported as relevant for resistive, ferroelectric, and spintronic devices \cite{youn2025ferroe,santra2025resist,castro2025spiket}, so integrating a physical array would require adding an analogue front end and validating against device-specific timing and compliance, rather than redesigning the controller. The shared interface allows the four blocks, or several instances of them, to be tiled into a larger array under one host. Wider neuron datapaths, multi-channel learning, and a larger crossbar address space are the immediate extensions, all of which fit the existing register model.

\section{Conclusion}\label{sec:concl}
This paper presented the design and verification of four interface-compatible neuromorphic blocks on the SkyWater 130\,nm process, spanning on-chip condition sensing, stochastic inference, local plasticity, and crossbar programming. The central outcome is that these functions, usually reported in isolation, can be built to one serial register model and a shared verification methodology, so that a single host can configure all four blocks through one access pattern while each block remains reusable on its own. Functional simulation establishes the intended behaviour of each block, and the open implementation flow places each block within a single tile at 61 to 70 per cent utilisation and meets the 50\,MHz target with positive setup and hold margin, with the designs submitted for tapeout to a shared-silicon (Tiny Tapeout) shuttle through a fully open digital integrated-circuit design flow, from register-transfer code to a manufacturable layout. Because the blocks are exercised per tile rather than as a fabricated system, the immediate next step is silicon measurement of frequency, energy, and array interaction, for which the observability built into each block is already provided. The shared interface and the register-defined crossbar controls give a concrete starting point for attaching memristive or spintronic arrays in later work.

\section*{Data and code availability}
The register-transfer source, test benches, OpenLane configuration, pinouts, and documentation for each of the four blocks are available in their respective public repositories: the ring-oscillator PVT sensor, \url{https://github.com/santhoshs93/tt_um_santhosh_ring_osc}; the stochastic neuron, \url{https://github.com/santhoshs93/tt_um_santhosh_stoch_neuron}; the STDP controller, \url{https://github.com/santhoshs93/tt_um_santhosh_stdp_ctrl}; and the crossbar controller, \url{https://github.com/santhoshs93/tt_um_santhosh_xbar_ctrl}. An index repository, \url{https://github.com/santhoshs93/tt_um_santhosh_neuromorphic_suite}, collects the four blocks as git submodules; it does not contain integrated top-level register-transfer code, a co-simulation, or a suite-level layout, which remain private design-level compositions. The results reported here correspond to the feature-complete revision of each block at the head of its default branch, namely commits \texttt{e99a4a3} (ring oscillator), \texttt{225ce6e} (stochastic neuron), \texttt{8385cf0} (STDP controller), and \texttt{718f170} (crossbar controller), all dated 20 April 2026; the figures and tables here should be reproduced at these commits.

\section*{Author contributions}
Santhosh Sivasubramani: Conceptualization, Methodology, Software (architecture and core implementation), Investigation, Validation, Writing (original draft), Writing (review and editing), Supervision, Project administration, Funding acquisition. Poornima Kumaresan: Data curation, Formal analysis, Visualization, Writing (review and editing).

\section*{Funding}
The authors acknowledge computational resources of the Intelligent Robotics and Rebooting Computing Chip Design (INTRINSIC) Laboratory, Centre for SeNSE, Indian Institute of Technology Delhi, and the IM00002G\_RB\_SG IoE Fund Grant (NFSG), Indian Institute of Technology Delhi.

\section*{Conflict of interest}
The authors declare no competing interests.

\section*{Acknowledgements}
The designs were implemented with the SkyWater 130\,nm process design kit and an open standard-cell flow, and were submitted through the Tiny Tapeout programme for shuttle access.

\clearpage
{\footnotesize
\bibliographystyle{ncestyle}
\bibliography{references}
}

\end{document}